\documentstyle[11pt,newpasp,twoside,epsf]{article}
\def\edcomment#1{\iffalse\marginpar{\raggedright\sl#1\/}\else\relax\fi}
\reversemarginpar

\begin{document}
\title{H- and He-burning central stars and the evolution to white dwarfs }
\author{T.\ Bl\"ocker}
\affil{Max--Planck--Institut f\"ur Radioastronomie, 53121 Bonn, Germany}

\begin{abstract}
The structure and evolution of central stars of planetary
nebulae (CSPNe) is reviewed.
CSPNe represent the rapid 
transitional stage between the Asymptotic Giant Branch (AGB)  
and the white-dwarf domain.
It is shown that the whole evolution off the 
AGB through the central-star regime depends on the evolutionary history. 
The detailed evolution into a white
dwarf is controlled by the internal stellar structure which, in turn, 
is determined 
by the duration of the preceding AGB evolution and therefore by the 
AGB mass-loss history. 
%
The evolution of hydrogen-deficient central stars has been a matter of debate
since many years. Convective overshoot appears to be a key ingredient to model 
these objects. Various thermal-pulse scenarios with inclusion of overshoot
are discussed, leading to surface abundances in general agreement with those 
observed for Wolf-Rayet central stars. 
\end{abstract}
\section{Introduction}
Central stars of planetary nebulae (CSPNe) represent the rapid 
transitional stage between the Asymptotic Giant Branch (AGB)  
and the white-dwarf domain. 
Since the pioneering work of Paczy\'{n}ski (1971), 
CSPNe have continued to be in the focus of evolutionary calculations.
For instance, 
Sch\"onberner (1979, 1983) 
demonstrated how the detailed evolution into a central star 
depends on the previous AGB evolution, i.e.\ on the thermal-pulse cycle
at the tip of the AGB and the shut down of the heavy AGB mass loss. 
The dependence of the CSPN evolution on the thermal-pulse cycle 
was investigated in full detail later by Iben (1984), 
and existing model grid calculations were complemented by the computations of 
Wood \& Faulkner (1986).
The next stage of calculations included the consideration of 
appropriate initial-final mass combinations
based on  empirical and semi-empirical AGB mass-loss 
prescriptions (Vassiliadis \& Wood 1993, 1994; Bl\"ocker 1995a,b). 
Further reviews on the AGB and post-AGB evolution 
and comparisons of the above calculations are given, e.g., by 
Iben (1995), Habing (1996), Wood (1997) and Sch\"onberner (1997).

While the evolution of hydrogen-rich central stars appears to be 
adequately understood, the formation of hydrogen-deficient central stars has 
been a matter of debate for many years.  Convective overshoot  
turned out to be a prerequisite to model the abundances of objects as the 
Wolf-Rayet central stars (Herwig et al.\ 1999, Bl\"ocker 2001, Herwig 2001a). 
 
\section{Moving off the AGB} 
Stars evolving along the AGB 
suffer from increasingly strong stellar winds with mass-loss rates
up to  $\sim 10^{-4}\,$M$_{\odot}$/yr (Habing 1996), 
eroding up to $\sim$80\% of the initial mass and terminating 
the AGB evolution when the envelope mass has dropped to 
$\approx 10^{-2}\,$M$_{\odot}$ (Sch\"onberner 1979). The  
stars move off the AGB, become central stars of planetary nebulae, 
and finally reach after exhaustion of 
nuclear burning the stage of white dwarfs. 
The complete evolution of a
3\,M$_{\odot}$ main-sequence star into a 0.625\,M$_{\odot}$ white dwarf
is illustrated in Fig.~1.

One the upper AGB, the He burning shell becomes recurrently unstable,
giving rise to the so-called thermal pulses 
(Schwarzschild \& H\"arm 1965, Weigert 1966), 
during which the luminosity of the He shell increases rapidly for a
short time of 100\,yr to $10^{6}...\hspace*{0.5mm} 10^{8}$\,L$_{\odot}$.
The huge amount of energy produced in the He shell
forces the development of a pulse-driven
convection zone which mixes products of He burning, i.e.\ carbon and oxygen,
into the intershell region.
Because the hydrogen shell is pushed concomitantly into cooler domains,
hydrogen burning ceases temporarily, allowing the envelope convection to 
proceed downwards after the pulse, 
to penetrate those intershell regions formerly enriched with
carbon (and oxygen), and to mix this material to the surface
(3$^{\rm rd}$ dredge up).
After the pulse, H burning re-ignites and provides again the main
source of energy.
%
Accordingly, the evolution off and beyond the AGB depends 
on the thermal-pulse cycle phase $\phi$
(fraction of the time span between two subsequent pulses), 
with which the stars move off the AGB. 
Fig.~2 shows the surface luminosity and the contributions of the H and 
He burning shell as a function of $\phi$ for the tenth pulse cycle of a 
3\,M$_{\odot}$ AGB sequence. 
The post-AGB evolution is dominated by He burning for $0 \le \phi \le 0.15$. 
For $0.15 \le \phi \le 0.3$ both nuclear shell sources contribute similar
luminosity fractions, for $0.3 \le \phi \le 1.0$ H~burning determines
the nuclear energy production 
(Iben 1984). If the thermal-pulse cycle-phase is sufficiently large 
($\phi > 0.75$), 
a last thermal pulse can occur
during the post-AGB evolution transforming a H burning into a He
burning  model.
The flash forces the star to expand rapidly to Red Giant dimensions, and the 
remnant quickly evolves back to the AGB (``born again scenario''). There, it  
starts its post-AGB evolution again, but now as a He burning object
(Iben 1984). 

%
\begin{figure}
\begin{minipage}{6.3cm}
\vspace*{1.5mm}
\epsfxsize=1.0\textwidth
\mbox{\epsffile{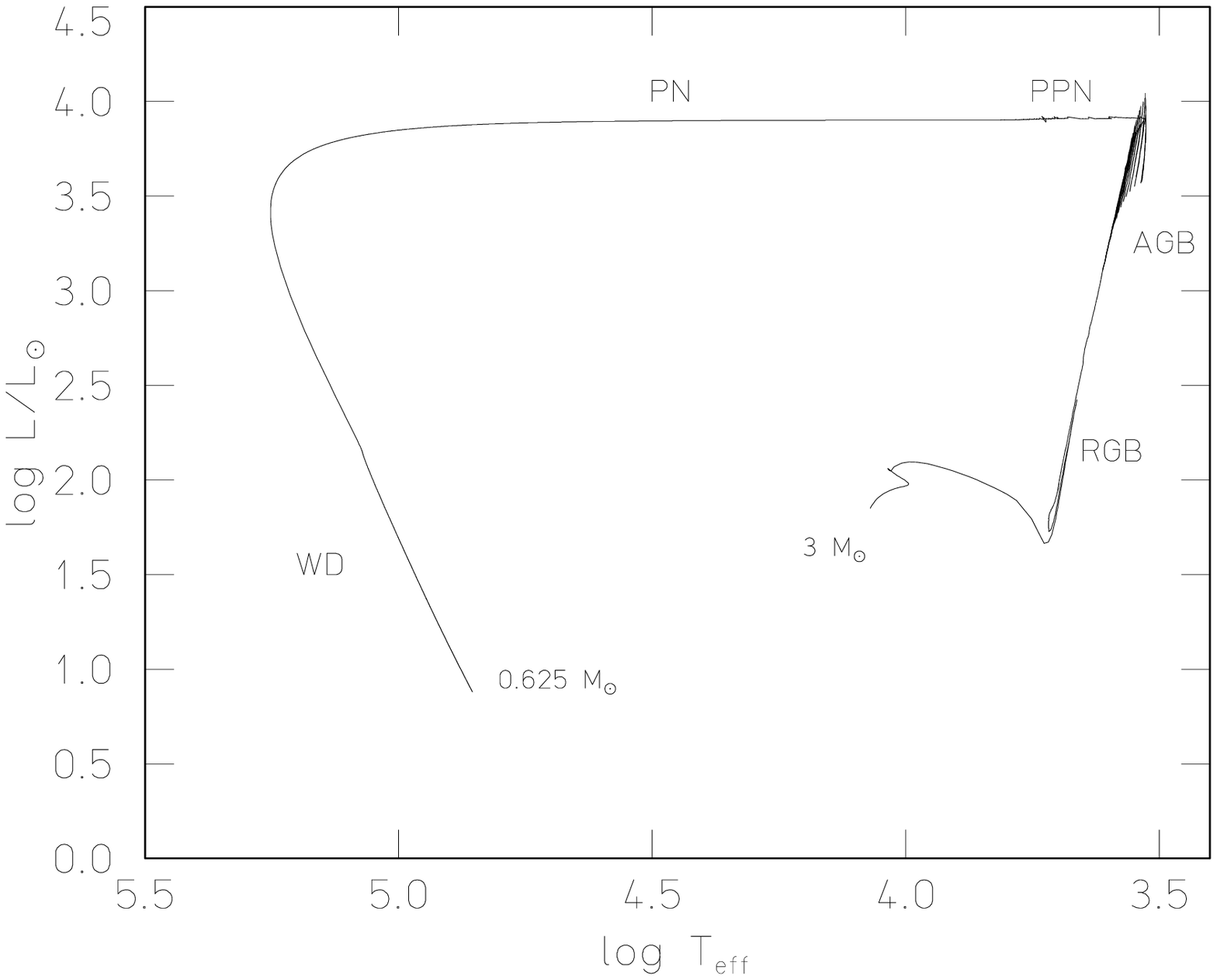}}
\end{minipage}
\begin{minipage}{6.3cm}
\hspace*{1.5cm}
\epsfxsize=0.83\textwidth
\mbox{\epsffile{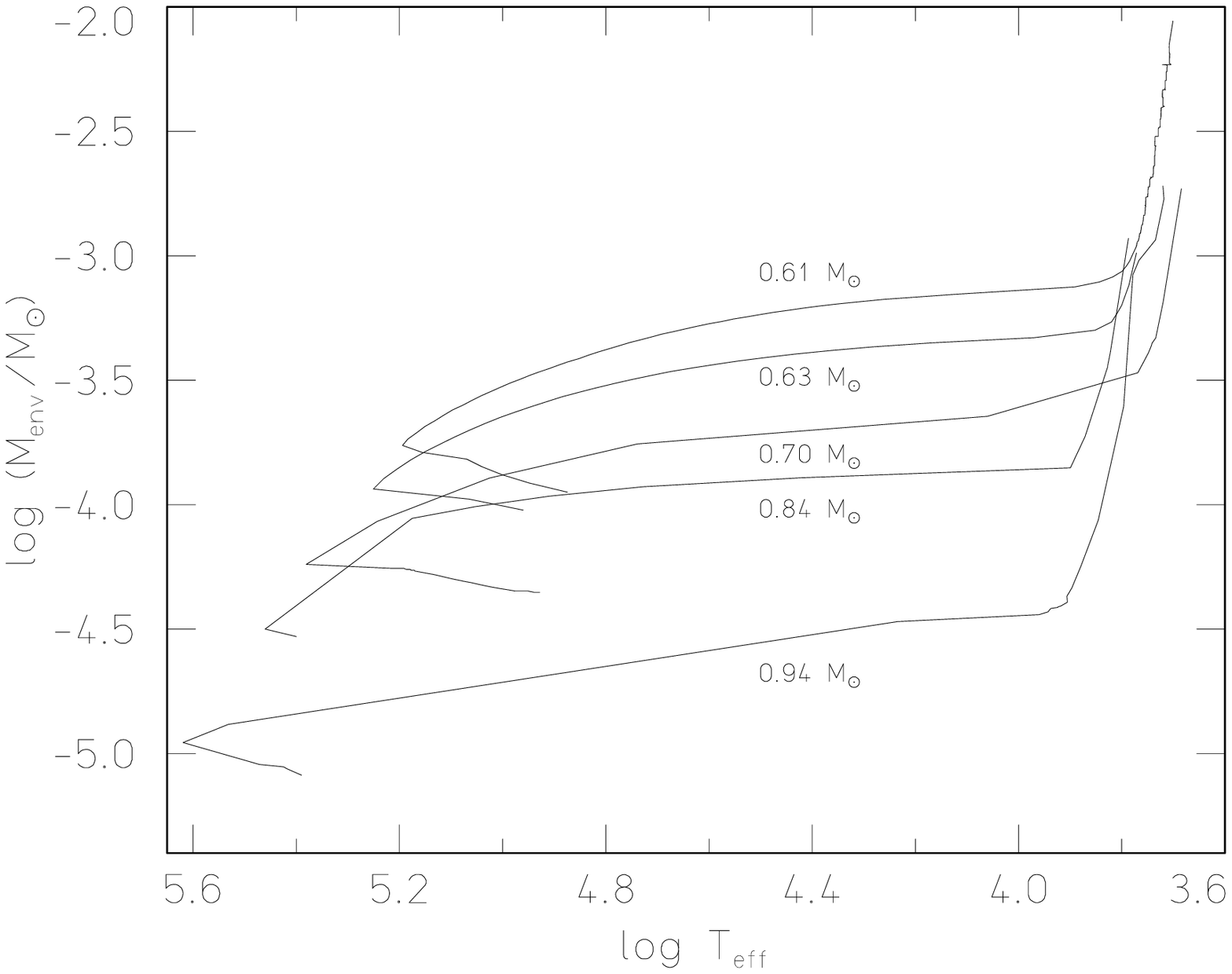}}
\end{minipage}
\caption{{\it Left:} Evolutionary path of a star with an initial mass of 
 3\,M$_{\odot}$ 
 from the main sequence through the AGB towards the white-dwarf stage. 
 On the AGB, the star suffers from 21 thermal pulses until mass loss 
 terminates the AGB evolution. The final mass is  0.625\,M$_{\odot}$. During
 the post-AGB evolution the model is burning hydrogen until extinction of 
 the shell source. 
 {\it Right:} Envelope mass vs.\ effective temperature for hydrogen burning
  post-AGB models (Bl\"ocker 1995b).
  The initial and remnant masses are
  (3\,M$_{\odot}$,0.61\,M$_{\odot}$),
  (3\,M$_{\odot}$,0.63\,M$_{\odot}$),
  (4\,M$_{\odot}$,0.70\,M$_{\odot}$),
  (5\,M$_{\odot}$,0.84\,M$_{\odot}$), and
  (7\,M$_{\odot}$,0.94\,M$_{\odot}$), resp.
}\label{Fmenvall}
\end{figure}
\begin{figure}
\hspace*{10mm}
\epsfxsize=0.85\textwidth
\epsfbox{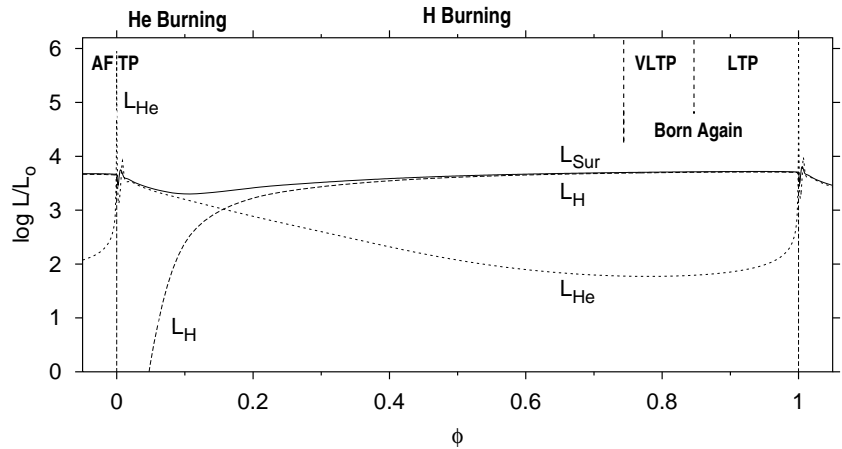}
\caption[phi]{Luminosity contributions due to H 
 ($L_{\rm H}$, dashed) and He burning 
 ($L_{\rm He}$, dotted) as well as surface lumininosity (solid) vs.\ 
 thermal-pulse phase $ \phi $ for the tenth
 pulse cycle of a 3 M$_{\odot}$ AGB sequence (Bl\"ocker 1995a).
 Critical phase ranges for the occurence of thermal pulses during (AFTP) 
 and after (LTP, VLTP) moving off the AGB are indicated  (see text).
} \label{Fphi}
\end{figure}

The details of the transformation of AGB stars into white dwarfs
is crucially determined by the treatment of mass loss on the AGB and beyond. 
Observations indicate that mass-loss should decrease by orders of magnitudes
during the transition to the central-star regime (Perinotto 1989).
However, at which temperature this
strong decrease takes place is only barely known.  
The transition time scale $t_{\rm H}$ of a H-burning model is determined 
by the shrinkage  of its envelope mass $ \Delta M_{\rm e}$ due to
burning  and mass loss, divided by the burning rate 
$\dot{M}_{\rm H}$ and the mass-loss rate $\dot{M}_{\rm w}$, i.e.\ 
$t_{\rm H} = \Delta M_{\rm e} / (\dot{M}_{\rm H} + \dot{M}_{\rm w})$
(Sch\"onberner 1983).
On the upper AGB, the evolution is controlled by mass loss whose rates 
are orders of magnitudes larger than those of burning 
($\dot{M_{H}} \sim 10^{-7} $M$_{\odot}$/yr for $M_{\rm H}$= $0.6 $M$_{\odot}$).
After shut down of the heavy AGB mass loss, burning takes over and mass loss
is only of minor importance. The earlier mass loss decreases, the longer are
the transition times, and the larger is the probability to obtain a final 
thermal pulse and thus a He burning model.
In the calculations of Vassiliadis \& Wood (1994),
the strong AGB mass-loss was switched off close to the AGB 
($T_{\rm eff}<5000$\,K) and the earlier the more 
massive the remnants are,
leading to transition times ($>10^{4}$\,yr) which increase with remnant mass.
Bl\"ocker (1995b) kept the high
AGB rate until the star's fundamental radial pulsation period has fallen 
below 50 days (including a short reduction phase; $T_{\rm eff}>5000$\,K). 
Accordingly, mass loss
shuts down the later the more massive the remnants are,
leading to a very rapid decrease of the transition times ($<10^{4}$\,yr) 
with remnant mass.  
The transition times should not be too long
since  the coolest post-AGB stars known have  
effective temperatures of about 5000\,K (Sch\"onberner 1997),
and kinematical ages of the youngest planetaries are only of the order of
1\,000 years.
Fig.~1 shows the envelope mass for different 
remnant masses as a function of the effective temperature. 

\section{From central stars to white dwarfs}
For the evolution through the PN region,
mass loss can be described by the radiation-driven wind theory 
(Pauldrach et al.\ 1988). An adaption yields
$ \dot{M}_{\rm CPN} = 1.3 \cdot 10^{-15} \, L^{1.9}$ (Bl\"ocker 1995b),
leading to rates of $10^{-8}$ to $10^{-7} $M$_{\odot}$/yr for remnants of
0.6 to 0.8\,M$_{\odot}$ being only of importance for massive remnants. 
Since $\Delta M_{\rm e}$ decreases and 
$\dot{M_{H}}$ increases with increasing core mass (see Fig.~1),
more massive H-burning remnants evolve faster than less massive ones 
along the horizontal part of the evolution.
Typical crossing times to evolve from $10^{4}$\,K 
to maximum effective temperature amount to 
$\sim 10^{5}$\,yr for 0.55 M$_{\odot}$, 
4000\,yr for 0.6 M$_{\odot}$, 800\,yr for  0.7 M$_{\odot}$, 
350\,yr for  0.84 M$_{\odot}$, 
and 50\,yr for 0.94 M$_{\odot}$.
The crossing timescale of He burning central stars formed in the 
born-again scenario is roughly three times larger. 

The ``plateau'' evolution at almost constant luminosity 
is terminated when the envelope mass becomes too 
small for hydrogen burning to be maintained. Then, the surface luminosity 
rapidly declines by more than one order of magnitude until it can be 
covered by gravothermal energy releases. He burning ceases as well in the 
course of evolution for most thermal-pulse phases. 
The further fading 
along the white dwarf cooling branch
is controlled by gravothermal energy releases and neutrino losses
and depends on the remnants' thermomechanical structure 
Though the evolutionary lines of degenerate cores
belonging to different initial masses do, in principle, converge in the 
density-temperature plane (Paczynksi 1970), implying that the fading into
white dwarfs depends only on the core mass, mass-loss terminates the 
AGB evolution well before such a convergence is reached (see Bl\"ocker 1995a).
Consequently, the thermomechanical structure and thus the fading times 
depend both on the core mass and the intial mass and thus on the complete 
evolutionary history. Only after sufficient cooling (e.g.\ $10^{6}$\,yr), 
long after dispersion of the planetary nebula, the evolution becomes 
more and more independent of the remnant's history.

The detailed fading into a white dwarf is intimately linked to the 
degeneracy of the core. During the AGB evolution the 
degeneracy of the core increases  and an increasing fraction of the energy
which is released by contraction is used up by raising the Fermi energy of the
electrons, being no longer available for the increase of the  thermal energy
of the star. Thus, the more degenerate a model leaves the AGB, 
i.e.\ the more compact and cooler the interior, the faster it fades after 
exhaustion of nuclear burning. 
For instance, since for a given initial mass the mean degeneracy increases 
with increasing AGB duration, models based on only one single initial mass
(Paczynski 1971, Wood \& Faulkner 1986)
fade the faster the more massive they are.
However, considering reasonable initial-final mass combinations,
more massive remnants can fade {\it slower} than less
massive ones (Bl\"ocker \& Sch\"onberner 1990, Bl\"ocker 1995b) since
their progenitors have 
a less compact and hotter interior than those of lighter ones.  
The dependence of the fading times on the (poorly known) AGB mass-loss 
history is underlined by a comparison with the models of
Vassiliadis \& Wood (1994), which are also consistent with initial-final mass 
relations. Their models dim faster than those of Bl\"ocker (1995b),
because they are based on AGB models which
spent a longer time on the AGB due to smaller mass-loss rates
and are thus more degenerate.   

Fig.~3 shows
the luminosity vs.\ remnant mass for both evolutionary sets
at a post-AGB age of $t$=$10^{4}$\,yr, the typical PN lifetime,
and illustrates that the least luminous central stars
should have masses of $\sim 0.65 $M$_{\odot}$ 
(Bl\"ocker \& Sch\"onberner 1990).  
The position of He-burning models is also shown in
Fig.~3. For lower masses, they evolve much slower than H-burning
models. For larger remnant masses, 
it is hard to distinguish between
the He-burning and the slowly fading H-burning model. 
Thus, in the Vassiliadis \& Wood  scenario 
massive and luminous central stars  
are most likely He-burning
objects whereas the calculations of Bl\"ocker (1995b) suggest 
that they can also be explained
by H-burning remnants.
\begin{figure}
\begin{minipage}{7.0cm}
\epsfxsize=1.0\textwidth
\epsfbox{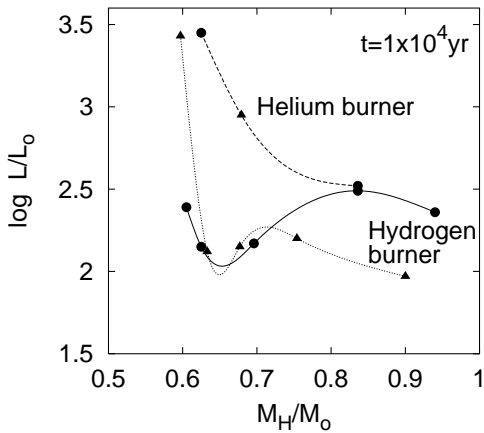}
\end{minipage}
\begin{minipage}{6.8cm}
\caption[iso]{Luminosity vs.\ core mass at $t = 10^{4}$\,yr for 
     H-burning (solid and dotted line) and He-burning models (dashed line) of
     Vassiliadis \& Wood (1994) [triangles] and Bl\"ocker (1995b) [circles].
} \label{Fiso}
\end{minipage}
\vspace*{-4mm}

\end{figure}
%
%
\vspace*{-6mm}

\section{Hydrogen-deficient central stars}
The origin and evolution of H-deficient post-AGB star
appeared to be enigmatic for many years.
Although stars
evolving through the AGB phase stay H-rich at their
surfaces, a considerable fraction of their descendants
show \hbox{H-deficient} compositions (Mendez 1991).
Approximately 20\% of the whole CSPNe population
seem to be H-deficient
while the rest show solar-like compositions.
Important constituents of the H-deficient population
are the Wolf-Rayet (WR) central stars and the hot
PG\,1159 stars with typical surface abundances of (He,C,O)=(33,50,17)
by mass (Dreizler \& Heber 1998, Koesterke \& Hamann 1997).
Standard stellar evolution calculations failed to model these objects
since they predict post-AGB stars 
either to have H-rich surfaces
(e.g.\ Bl\"ocker \& Sch\"onberner 1997) or, if H-deficient,
to expose only a few percent of oxygen in their photospheres
(Iben \& McDonald 1995).
However, if convective overshoot is considered,
H-deficient post-AGB stars with
abundance patterns as observed in WR central stars can be formed
(Herwig et al.\,1999, Bl\"ocker\,2001, Herwig\,2001a).

\begin{figure}
\hspace*{-2mm}
\epsfxsize=0.95\textwidth
\epsfbox{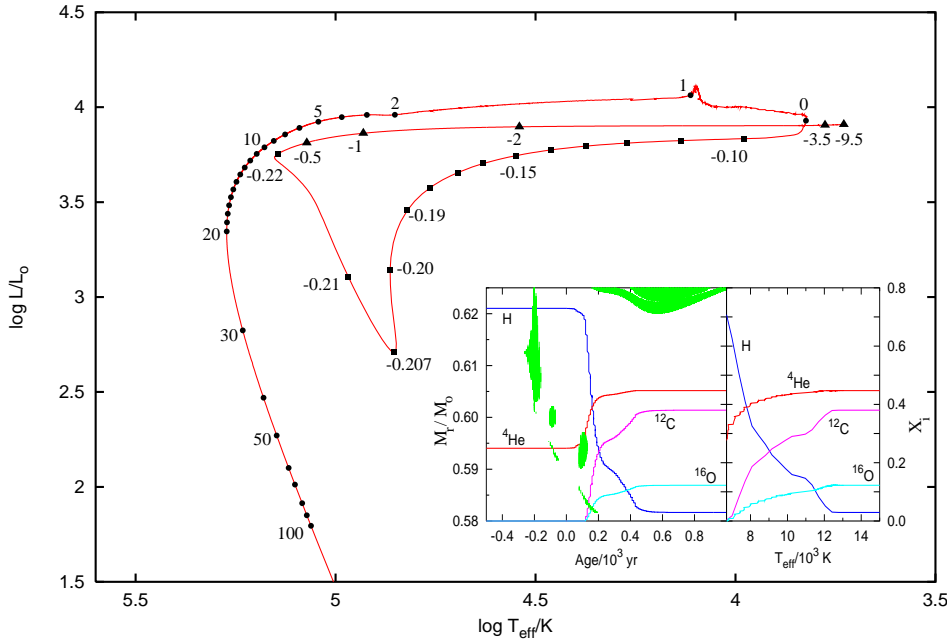}
\vspace*{-6mm}
\caption[track]{
Evolutionary track for a  0.625\,M$_{\odot}$ overshoot model suffering from an
LTP (Bl\"ocker 2001).
Time marks are given in units of $10^{3}$\,yr.
Time is set to zero at minimum effective temperature after the flash.
The age -3500\,yr marks the begin of the central-star evolution and refers to
a pulsational period of 50\,d. The inlet shows:
{\it Left:}  
Extension of convective regions (left scale, shaded) and
surface abundances of H, He, C, and O (right scale) vs.\ time
during flash and  dredge up. {\it Right:} 
The surface abundances 
vs.\,$T_{\rm eff}$ for the dredge-up phase.}
\end{figure}
Herwig et al.\ (1997) showed that diffusive overshoot
(Freytag et al.\ 1996) applied to all convective regions 
leads for AGB models to 
(i) efficient dredge-up and formation of low-mass carbon
stars;
(ii) formation of $^{13}$C as a neutron source to drive $s$-process 
nucleosynthesis; and
(iii) considerable changes of the intershell abundances 
The latter finding (iii) turned out to be a key ingredient for the modelling of
WR central stars. Overshoot leads
to an enlargement of the pulse-driven convection zone and
to enhanced mixing of core material from deep layers below the He shell
to the intershell zone
(``intershell dredge-up'')
resulting in intershell abundances (mass fractions) of (He,C,O)=(40,40,16)
instead of (70,25,2) as in non-overshoot sequences.
These modified intershell abundances are already close to the 
observed surface abundances of Wolf-Rayet central stars.
Finally, in contrast to standard evolutionary calculations, 
overshoot models do show dredge up for very low envelope masses, 
and efficient dredge up was found even during the post-AGB 
stage (Bl\"ocker 2001, Herwig 2001a) leading to the mixing of the intershell
abundances to the surface and to the dilution of hydrogen. 
Three thermal pulse scenarios for Wolf-Rayet central stars can now
be distinguished:  \\
{\bf AGB Final Thermal Pulse (AFTP)}, 
occurring immediately before the star moves off the AGB 
In this case the envelope mass is already very small
($\sim 10^{-2} $M$_{\odot}$, Fig.~1). During dredge-up
a substantial fraction of the intershell region is mixed with the
tiny envelope leading to the dilution of hydrogen and enrichment
with carbon and oxygen. The resulting surface abundances
depend on the actual envelope mass at which the AFTP occurs.
For instance, 
Herwig (2001a) found for M$_{\rm env}= 4 \cdot 10^{-3} $M$_{\odot}$ 
(H,He,C,O)=(17,33,32,15) 
after an AFTP.
To obtain a sufficiently high probability 
for the AFTP ($\phi \approx 0$, Fig.~2)
requires, however, 
a coupling of mass loss to the thermal pulse cycle.
The AFTP leads to a relatively high hydrogen abundance ($\ga 15\%$)
and predicts small kinematical ages for the PNe of WR central stars which 
emerge here directly from the AGB. \\
{\bf Late Thermal Pulse (LTP)},
occurring when the model evolves
with roughly constant luminosity from the AGB towards the white dwarf
domain ($\phi \ga 0.85$, Fig.~2).
This kind of thermal pulse is similar to an AFTP but the envelope 
mass is even smaller ($\sim 10^{-4} $M$_{\odot}$, Fig.~1).
Fig.~4 shows the evolutionary track of a 0.625\,M$_{\odot}$
LTP model with diffusive overshoot (Bl\"ocker 2001).
After the flash the intershell abundances amount to (He,C,O)=(45,40,13) and
the model evolves towards the AGB domain on a timescale of $\sim 100$\,yr. 
At minimum effective temperature ($\approx 6700$\,K) dredge up sets in 
and continues until the star has reheated to $\approx 12000$\,K.
Hydrogen is diluted to 3\% and the final surface abundances of He, C and O 
are close to those of the intershell region, viz.\ (45,38,12). Extension of 
convective regions and abundances are illustrated in the inlet of 
Fig.~4 as 
function of age and effective temperature,
resp. The kinematical age of the PN amounts to a few thousand years. 
An observed example of an LTP is the born-again object FG\,Sge 
(Bl\"ocker \& Sch\"onberner 1997, Bl\"ocker 2001).  
\\ 
{\bf Very  Late Thermal Pulse (VLTP)},
occurring when the model is already on the  white dwarf cooling track,
i.e.\ after the cessation of  H burning 
($0.75 \la \phi \la 0.85$, Fig.~2).
Then, the pulse-driven convection zone can reach and penetrate the H-rich
envelope and protons are ingested into the hot, carbon-rich intershell 
region 
raising a H flash (Fujimoto 1977, Sch\"onberner 1979, Iben \& McDonald 1995).
The energy released by this flash leads to a splitting of the convection 
into an upper zone powered by H burning and a lower one powered by He burning. 
The upper convection zone is, however, short-lived because the available 
hydrogen in the envelope is quickly consumed. Finally, the star becomes 
hydrogen-free and exposes its intershell abundances at the surface. 
Herwig et al.\ (1999) found for a 0.604\,M$_{\odot}$ overshoot model
surface abundances of (He,C,O)=(38,36,17). 
The kinematical age of the PN is relatively high
since the star has first to fade along the cooling branch down
to a few 100\,L$_{\odot}$ before the flash sets in. 
For 0.6 M$_{\odot}$ one obtains typically $t \ga  20000$\,yr.
The central star V\,4334\,Sgr is thought to have recently suffered from a VLTP
(Herwig 2001b).

All scenarios lead to hydrogen-deficient post-AGB stars with 
carbon and oxygen abundances as observed for Wolf-Rayet central stars.
The variety of observations requires most likely all of these scenarios.
Many objects have only very low hydrogen abundances, if any, favoring the 
LTP and VLTP. On the other hand, several Wolf-Rayet central stars are 
surrounded by young  planetary nebulae (Tylenda 1996) and circumstellar shells 
consisting of both C- and O-rich material 
(Waters et al.\ 1998, Cohen et al. 1999) strengthening the AFTP and LTP. 
Within the current models roughly 20 to 25\% of the stars moving off the AGB 
can be expected to become hydrogen-deficient.

\end{document}